\documentclass[conference]{IEEEtran}
\usepackage{cite}
\ifCLASSINFOpdf
\usepackage[pdftex]{graphicx}
\else
\usepackage[dvips]{graphicx}
\fi
\usepackage{amsmath}
\usepackage{array}
\usepackage{url}
\usepackage{color}
\usepackage{txfonts}

\usepackage{listings}

\usepackage{booktabs}
\begin{document}
%
% paper title
% Titles are generally capitalized except for words such as a, an, and, as,
% at, but, by, for, in, nor, of, on, or, the, to and up, which are usually
% not capitalized unless they are the first or last word of the title.
% Linebreaks \\ can be used within to get better formatting as desired.
% Do not put math or special symbols in the title.
\title{Log-based Anomaly Detection\\ of CPS %System
Using a Statistical Method}
%\title{A Log-based Anomaly Detection of a CPS}

% author names and affiliations
% use a multiple column layout for up to three different
% % affiliations
% \author{\IEEEauthorblockN{Yoshiyuki Harada}
% \IEEEauthorblockA{Department of Information Science\\
% Kyoto Institute of Technology\\
% Matsugasaki Goshokaido cho, Sakyo-ku,\\
% Kyoto 606-8585, Japan
% Email:}
% \and
% \IEEEauthorblockN{Yoriyuki Yamagata}
% \IEEEauthorblockA{National Institute of Advanced \\
% Industrial Science and Technology (AIST)\\
% 1-8-31 Midorigaoka, Ikeda,\\
% Osaka 563-8577, Japan\\
% Email: yoriyuki.yamagata@aist.go.jp}
% \and
% \IEEEauthorblockN{Osamu Mizuno}
% \IEEEauthorblockA{Department of Information Science\\
% Kyoto Institute of Technology\\
% Matsugasaki Goshokaido cho, Sakyo-ku,\\
% Kyoto 606-8585, Japan
% Email: o-mizuno@kit.ac.jp}
% \and
% \IEEEauthorblockN{Eunhye Choi}
% \IEEEauthorblockA{National Institute of Advanced\\
% Industrial Science and Technology (AIST)\\
% 1-8-31 Midorigaoka, Ikeda,\\
% Osaka 563-8577, Japan\\}
%}

% conference papers do not typically use \thanks and this command
% is locked out in conference mode. If really needed, such as for
% the acknowledgment of grants, issue a \IEEEoverridecommandlockouts
% after \documentclass

% for over three affiliations, or if they all won't fit within the width
% of the page, use this alternative format:
%
\author{\IEEEauthorblockN{Yoshiyuki Harada\IEEEauthorrefmark{1}\IEEEauthorrefmark{2},
Yoriyuki Yamagata\IEEEauthorrefmark{2},
Osamu Mizuno\IEEEauthorrefmark{1},
Eun-Hye Choi\IEEEauthorrefmark{2}}
\IEEEauthorblockA{\IEEEauthorrefmark{1}
%Department of Information Science\\
Kyoto Institute of Technology, Kyoto, Japan\\
%Matsugasaki Goshokaido cho, Sakyo-ku,\\
%Kyoto 606-8585, Japan
Email: y-harada@se.is.kit.ac.jp, o-mizuno@kit.ac.jp}
\IEEEauthorblockA{\IEEEauthorrefmark{2}National Institute of Advanced
Industrial Science and Technology (AIST), Ikeda, Japan\\
%1-8-31 Midorigaoka, Ikeda,\\
%Osaka 563-8577, Japan\\
Email: \{yoriyuki.yamagata, e.choi\}@aist.go.jp}}

% use for special paper notices
%\IEEEspecialpapernotice{(Invited Paper)}

% make the title area
\maketitle

% As a general rule, do not put math, special symbols or citations
% in the abstract
\begin{abstract}
Detecting anomalies of a \emph{cyber physical system} (\emph{CPS}), which is a complex system consisting of both physical and software parts, is important because a CPS often operates autonomously in an unpredictable environment.
However, because of the ever-changing nature and lack of a precise model for a CPS, detecting anomalies is still a challenging task.
To address this problem, we propose applying an \emph{outlier detection} method to a CPS log.
By using a log obtained from an actual aquarium management system, we evaluated the effectiveness of our proposed method by analyzing outliers that it detected.
By investigating the outliers with the developer of the system, we confirmed that some outliers indicate actual faults in the system.
For example, our method detected failures of mutual exclusion in the control system that were unknown to the developer.
Our method also detected transient losses of functionalities and unexpected reboots.
On the other hand, our method did not detect anomalies that were too many and similar.
In addition, our method reported rare but unproblematic concurrent combinations of operations as anomalies.
Thus, our approach is effective at finding anomalies, but there is still room for improvement.
\end{abstract}

% no keywords

% For peer review papers, you can put extra information on the cover
% page as needed:
% \ifCLASSOPTIONpeerreview
% \begin{center} \bfseries EDICS Category: 3-BBND \end{center}
% \fi
%
% For peerreview papers, this IEEEtran command inserts a page break and
% creates the second title. It will be ignored for other modes.
\IEEEpeerreviewmaketitle

\section{Introduction}

A \emph{cyber physical system} (\emph{CPS}) is a complex system consisting of both physical and software parts.
CPSs have become increasingly ubiquitous in our lives with the advent of machines controlled by software, such as drones, driverless cars, and automatically operated trains.
Nowadays, not only are more machines operated automatically, but those machines are also connected by a network and make collective decisions without human intervention.

Our research goal is to find \emph{anomalies} in a CPS by fully automated means.
Anomaly detection is especially important for a CPS because anomalies in the autonomous operation of physical objects in the CPS can cause losses of not only property but also even human life~\cite{Gurdian}.
A CPS operates in an ever-changing environment in an ever-changing way.
This makes testing a CPS difficult or impossible before the system is deployed.
Moreover, a lack of precise models for both a CPS and its environment makes formal analysis burdensome.
Therefore, we need a method to detect CPS anomalies in real-time or post mortem and correct the problem.

In this paper, we applied an \emph{outlier} detection technique for CPS anomaly detection.
Outlier detection techniques have been extensively studied in the fields of statistics, machine learning, and data mining.
An outlier is a data point that is suspected to be generated by a different mechanism than most data points.
An outlier is not necessarily an anomaly, but we focused on them because the latter has a good chance to be the former.
We applied the \emph{Local Outlier Factor (LOF)}~\cite{Breunig2000}, which is widely used for outlier detection of multi-dimensional real-value data, to a log of an automatic aquarium management system and evaluated its effectiveness at anomaly detection.
An advantage of the LOF is that no assumption is necessary regarding the distribution of data points.

However, there are several challenges to applying the LOF to CPS anomaly detection.
First, a CPS log is a mixture of discrete and real-value data.
The LOF is not designed for discrete data.
We resolved this problem by mapping log entries to real-value vectors in a high-dimensional space.
This mapping is straightforward.
Second, the LOF is not designed for a time series.
We resolved this problem by concatenating vectorized log entries in a moving window.
In this manner, we can detect \emph{position} and \emph{combination} outliers, which are necessary to detect CPS anomalies.

For evaluation, we used an automatic aquarium management system called \emph{Aqua-tan}~\cite{aquatan} as a scaled example of a CPS.
For experiments, we collected 1,000,000 log entries of Aqua-tan and separated them into 10 chunks.
For each chunk, we applied the proposed method.
The ELKI tool~\cite{DBLP:journals/pvldb/SchubertKEZSZ15} was used to calculate the \emph{outlier factor} for the LOF algorithm.
Finally, we analyzed five log sequences that achieved the highest outlier factors for each chunk by comparing them with a cluster of nearby data points.

Through this analysis, we evaluated the effectiveness of our method at finding CPS anomalies.
Our method detected several interesting anomalies.
For example, our method detected failures of mutual exclusion in the control system that were unknown to the developer.
Our method also detected anomalies such as temporal losses of functionalities, unexpected reboots, and manual interventions to the system.
On the other hand, we found several limitations for our method.
For example, if there are too many anomalies in sequence, our method detected a sequence of anomalies as inliers.
In addition, our method reported rare but unproblematic concurrent combinations of operations as outliers.
We concluded that our approach is effective at finding faulty anomalies, but there is still room for improvement.
\section{Related Work}
\label{sec:rw}

Anomaly detection is a research field that uses diverse methods for diverse applications.
For details, please refer to the survey~\cite{Chandola2009} or book~\cite{Aggarwal:2015:OA:2842756}.
Because the literature on anomaly detection is very extensive, we describe only the work relevant to the CPS, anomaly detection from a software log, and alternative methods for LOF here.

There is extensive literature on anomaly detection of a hybrid system~\cite{Verma2004,Narasimhan2007,Hofbaur2002,Zhao2005,Hofbaur2004};
all of them presuppose a model of a system.
On the other hand, our method does not assume a system model because preparing the model for CPS and its environment is a difficult task.

There is also extensive literature on anomaly detection from a software log~\cite{Russo2014,He2016,Xu2009,Chen}. These papers either assume that a log is purely discrete or real-value data.
On the other hand, our method handles a log with both discrete and real-value data.

We handled vectors with very high dimensions in our experiment; thus, we pushed the outlier detection method to its limit.
Outlier detection in high-dimensional space is an active research area~\cite{Aggarwal2013}.
In addition to LOF, we attempted to use high contrast subspaces (HiCS) algorithm ~\cite{Keller2012} and the correlation outlier probability (COP) algorithm~\cite{Kriegel2012}; however, their computations could not be completed within reasonable time and memory constraints.

\section{Preliminary: Local Outlier Factor (LOF)}
\label{sec:LOF}
In this section, we introduce the LOF algorithm based on the work of Aggarwal~\cite{Aggarwal:2015:OA:2842756}.
The LOF algorithm uses a fixed $k \geq 1$ to determine the \emph{outlier factors} of data points.
For each data point $x$, the outlier factor is determined as follows.
Let $D_k(x)$ be the distance of the $k$-nearest neighbor of $x$ and $L_k(x)$ be the data points within $D_k(x)$ from $x$.
Now, the \emph{reachability distance} $R_k(x, y)$ from $x$ to a data point $y$ is defined as

\begin{equation}
R_k(x, y) = \max \{d(x, y), D_k(y) \}
\end{equation}
where $d(x, y)$ is the distance between $x$ and $y$.
Typically, the Euclidean distance is used for $d(x, y)$.
Then, the \emph{average reachability distance} $AR_k(x)$ is defined as the average of the reachability distance from $x$ to data points in $L_k(x)$:

\begin{equation}
AR_k(x) = \frac{1}{\#L_k(x)} \sum_{y \in L_k(x)}R_k(x, y)
\end{equation}
where $\#L_k(x)$ is the number of data points in $L_k(x)$.
$\#L_k(x)$ may not be equal to $k$ because of a tie.
Finally, the outlier factor of $x$ is defined as the average of the ratio between $AR_k(x)$ and $AR_k(y), y \in L_k(x)$:

\begin{equation}
LOF_k(x) = \frac{1}{\#L_k(x)} \sum_{y \in L_k(x)} \frac{AR_k(x)}{AR_k(y)}
\end{equation}
If the outlier factor for $x$ is large, $x$ is more likely to be an outlier.

\section{Target System and Proposed Method}
\label{sec:method}

\begin{figure}[tbp]
\centering
\includegraphics[width=1.0\linewidth]{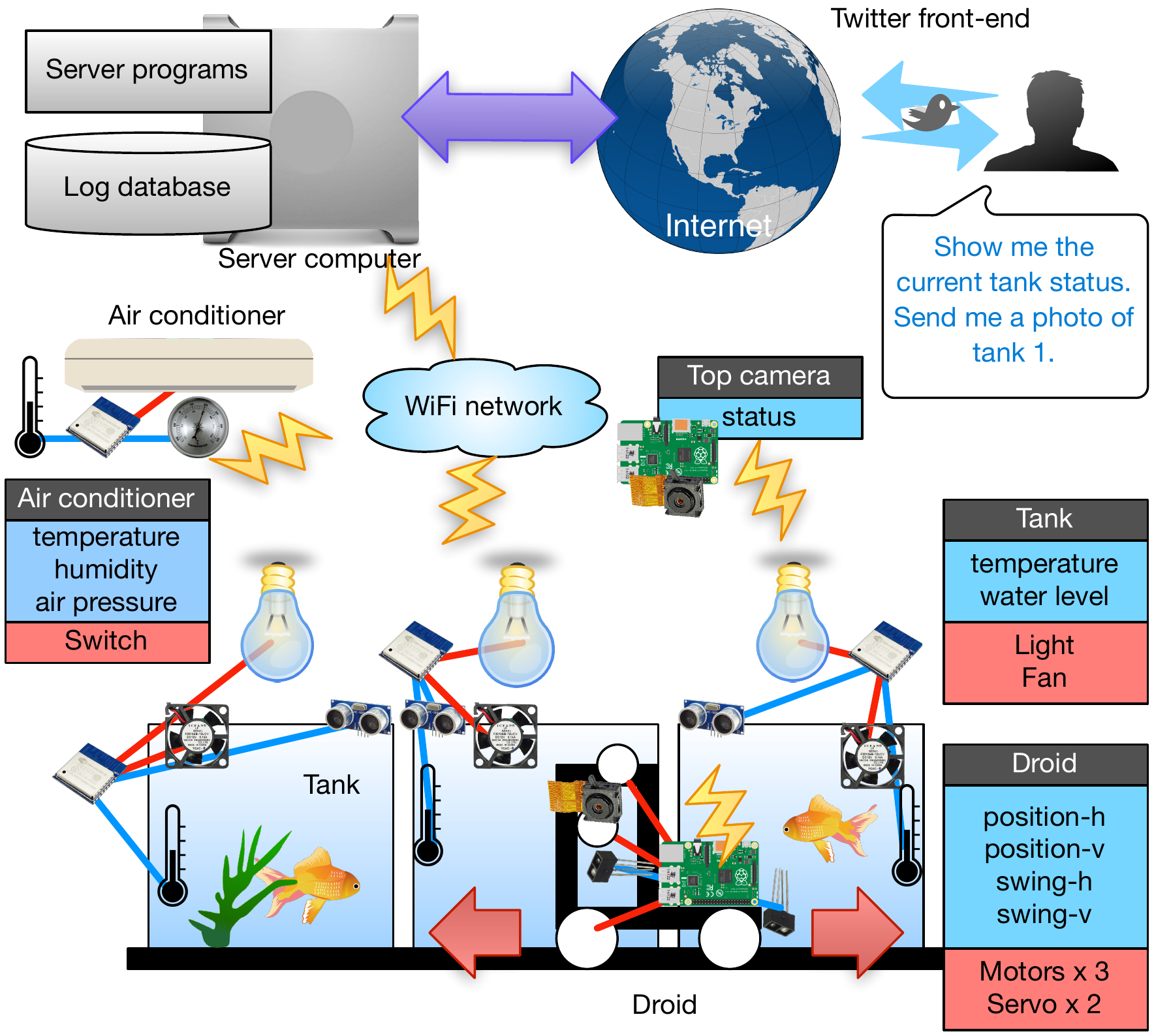}
\caption{Aqua-tan: automatic aquarium management system.}
\label{fig:aquatan}
\end{figure}

\subsection{Target System: Aqua-tan}

Our target system is the automatic aquarium management system \emph{Aqua-tan}\footnote{\url{https://se.is.kit.ac.jp/aquarium/}}~\cite{aquatan}.
Aqua-tan is a CPS for aquarium management with sensors and actuators connected to a WiFi network.
Aqua-tan mainly consists of three units: tanks, a droid, and air conditioner control.
The tank units monitor thermometers for the water temperature and a supersonic sensor for the water level.
They also control the lighting and fans to manage the environment of the aquarium.
The droid unit moves between tanks, takes photos of tanks from various positions and angles, and automatically feeds all tanks.
The air conditioner control unit switches the air conditioner on or off according to the air temperature of the room.
Several other units provide additional functionalities, such as the top view camera server.
These units can be controlled by the Twitter front-end, which accepts orders for the tank and droid units in natural language.
The Twitter front-end also provides periodic information regarding the tank status and photos.

Aqua-tan generates a log entry for each system command; an example is shown in Fig.~\ref{src:logexample}.
Each log entry consists of five columns: the ID, a time stamp, a command name, numerical data, and character string data.
The ID is a count of log entries from the beginning.
The time stamp is the time when the log entry is recorded.
The command name is generated by the system.
Aqua-tan has 130 commands that are subdivided into the ``Actuator drive'', ``Sensor value'', ``Network status'', and ``Others'' (Table \ref{tbl:aquatan_commands}).
Each command has either a numerical argument or string data (name).
For example, the command ``\verb|humidity|'', which shows the humidity of the room, has a numerical argument that indicates the humidity measurement of 30, as shown in Fig.~\ref{src:logexample}.
On the other hand, the command ``\verb|fan1_status|'' does not have a numerical argument but the string argument ``on'', which indicates that fan 1 is on.

All log entries are stored in a MySQL database.
We use the CSV format for the log entries extracted from this database.

\begin{table}[]
\centering
\caption{Aqua-tan commands.}
\label{tbl:aquatan_commands}
\begin{tabular}{lrl}
\toprule
Class & Kinds & Example \\
\midrule
Actuator drive & 46 & droid\_tank\_pos, droid\_lift\_pos, $\ldots$ \\
Sensor value & 31 & cputemp, humidity, water1, $\ldots$ \\
Network status & 7 & target\_X.X.X.X\_status, $\ldots$ \\
Others (Exclude) & 46 & fridge1\_status, location\_X, $\ldots$ \\
\bottomrule
\end{tabular}
\end{table}

\begin{figure}[tbp]
\begin{lstlisting}
39993,"2014-06-06 22:00","air",28,NULL
39994,"2014-06-06 22:00","humidity",30,NULL
39995,"2014-06-06 22:06:18","fan1_status",NULL,"on"
39996,"2014-06-06 22:10","cputemp",49.8,NULL
39997,"2014-06-06 22:10","pressure",994.1,NULL
39998,"2014-06-06 22:10","water2",27.1,NULL
39999,"2014-06-06 22:10","water1",26,NULL
40000,"2014-06-06 22:10","water3",27.8,NULL
40001,"2014-06-06 22:10","air",28,NULL
40002,"2014-06-06 22:10","humidity",30,NULL
\end{lstlisting}
\caption{Example Aqua-tan log.}\label{src:logexample}
\end{figure}

\subsection{Proposed Method}\label{subsec:method}

Our method for detecting anomalies of the target system consists of six steps.
\begin{enumerate}
\item \textbf{Pre-processing}: Collect log entries related to the automatic management of the aquarium.
\item \textbf{Vectorization}: Convert each log entry to a real-value vector.
\item \textbf{Normalization}: Normalize each vector component so that each component has 0 as the average and 1 as the variance.
\item \textbf{Windowing}: Concatenate the vectors in the moving window to create a single vector.
\item \textbf{LOF}: Calculate the outlier factor for each window using the LOF algorithm.
\item \textbf{Post-processing}: Take entries of high outlier factors as anomalies.
\end{enumerate}

\textbf{Pre-processing}. First, we collect log entries that are related to the targeted system.
Aqua-tan logs contain entries that are related to the status of mobile terminals and results of sentiment analysis on Twitter statements.
These entries are removed by pre-processing because they cause spurious outliers.

\textbf{Vectorization}. Next, each log entry is converted to a real-value vector.
The first component of the vector is the time (in milliseconds) that has passed from the previous entry of the same command.
The rest of the vector components are determined as follows.
For each command in the log entries, we prepare a vector $v$ of the five components that are filled by the real value 0.
If a command has the numerical argument $a$, we put $a$ into the first component $v[0] = a$.
If a command has the string argument $s$, we first take the hash value $h(s)$ of its string argument and compute the remainder $h(s) \% 4$ of the hash value when divided by 4.
Then, we put $v[1+h(s) \% 4] = 1$.
Finally, all vectors for all commands are concatenated in the predetermined fixed order.
Note that only one component is non-zero in the vector that is obtained by this procedure, excluding the first component.

\textbf{Normalization}. Third, we normalize vectors by linearly transforming each vector component, to have 0 as the average and 1 as the variance.

\textbf{Windowing}. Fourth, each vector that is obtained with the above method is concatenated by using a moving window of a fixed size.
This creates vectors that reflect combinations of commands and their arguments of successive time windows.

\textbf{LOF}. Fifth, the vectors are fed to the LOF algorithm, and outlier factors are computed.

\textbf{Post-processing}. Finally, we consider log entries that belong to windows with high outlier factors as anomalies.

\section{Experiments and Analysis}
\label{sec:exp}

\subsection{Experimental Setup}
In our experiment using Aqua-tan log entries, we first created 10 chunks consisting of 100,000 consecutive log entries from the entire log database.
We created chunks because the log spanned 2.2 years, and the system changed significantly during this period.
Therefore, there was not much sense in processing the entire log at once.
Another reason was that the LOF algorithm has a computation time of $O(n \log n) \sim O(n^2)$ depending on the dimension $n$ of the data.
Thus, chunking reduces the computation time.

\begin{figure}[tbp]
\centering
\includegraphics[width=1.0\linewidth]{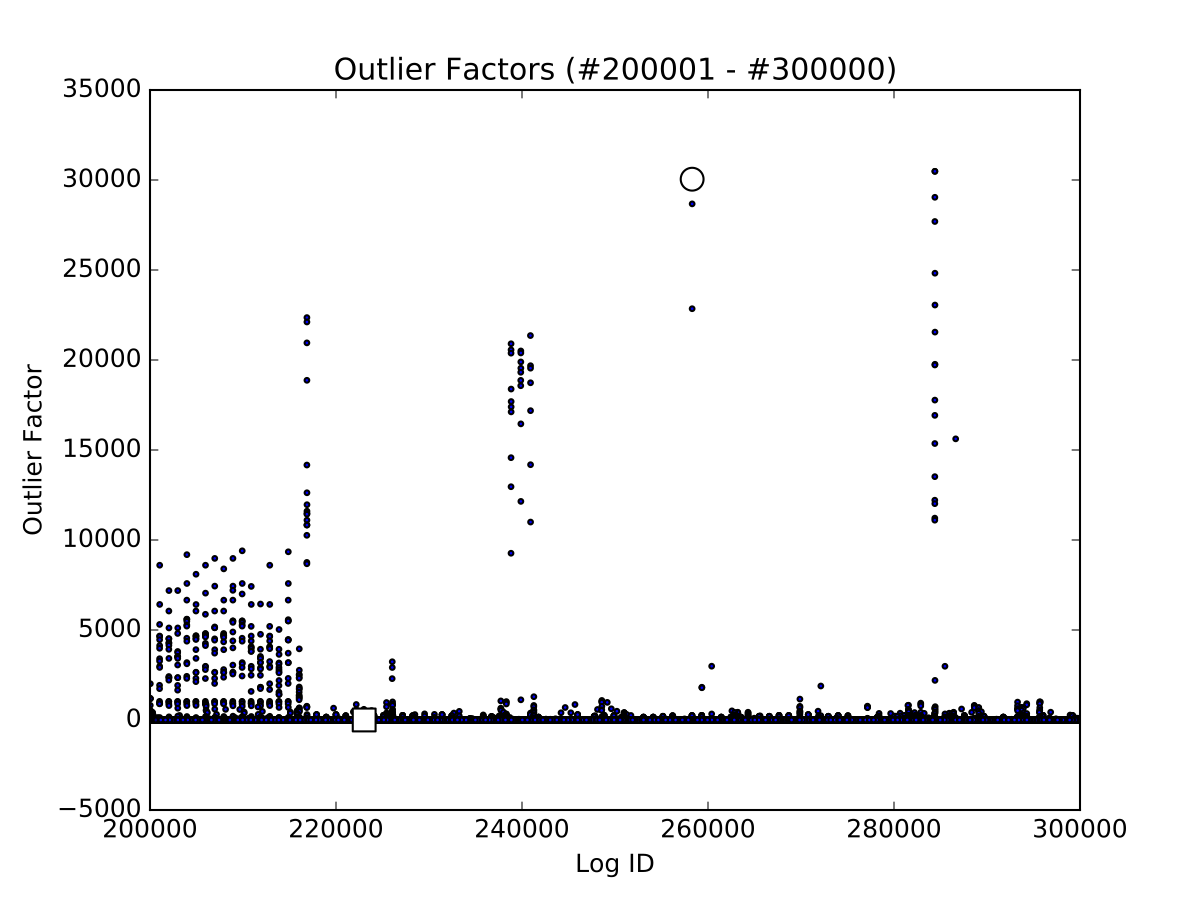}
\caption{Time series of outlier factors: from \#200000 to \#300000.}
\label{fig:lof_graph}
\end{figure}

Each chunk was processed independently.
Pre-processing, vectorization, normalization, and the taking of a moving window were all done with Python scripts.
We vectorized each log entry into 191--301-dimension vector space.
The dimension of the vectorized log entry was determined from the number of commands.
For example, chunk1 had 196 dimensions, while chunk4 had 201 dimensions.
In the windowing phase, we set the window size to 11.
The vector concatenated with a window width of 11 had about 2101--3311 dimensional vectors for each window.
Then, outlier factors were computed with the standalone tool ELKI~\cite{DBLP:journals/pvldb/SchubertKEZSZ15}.
We set $k = 20$ based on the recommendation by Breunig et al.~\cite{Breunig2000}.
Fig.~\ref{fig:lof_graph} visualizes the ELKI output.
The circle marker indicates an example from Table \ref{fig:ex_outliers}(a), and the square marker indicates an example from Table \ref{fig:ex_outliers}(b).
Finally, the output of ELKI was post-processed with Python scripts.
During the post-processing, five windows of the highest outlier factors were extracted, and $k = 20$ log sequences (i.e., $k$-neighborhoods) that were near the outlier were also recorded for later analysis.
% In Python scripts, we used the Python libraries pandas~\cite{pandas:PythonDataAnalysisLibrary} for processing CSV files and NumPy~\cite{NumPy:NumPy} for normalizing vectorized data and calculate the Euclidean distance between two vectors.

For computing, we used a machine with one Intel(R) Xeon(R) CPU E5-1620 v3, 3.50GHz processor having four cores and 64 GB RAM.
It took about 30 hours to complete all tasks of the proposed method.

\subsection{Analysis}

For analysis, we classified the extracted 50 log sequences of outliers by the cause of every outlier.
Table \ref{tbl:outlier_clusters} summarizes the classification results: seven causes of outliers and the number of outliers in each category.
In the following, we explain our analysis results for each category.
Example outliers for categories 1--5 are shown in Fig.~\ref{fig:ex_outliers}; because of space limitations, we omit examples for categories 6 and 7 in the figure.

\begin{table}[tb]
\centering
\caption{Outlier classification.}
\label{tbl:outlier_clusters}
\begin{tabular}{cp{6.5cm}r}
\toprule
& Cause of outlier & Count \\
\midrule
1 & Failure of mutual exclusion & 2 \\
\cmidrule{1-3}
2 & Unexpected reboot & 4 \\
\cmidrule{1-3}
3 & Failure of single functionality & 6 \\
\cmidrule{1-3}
4 & Mixture of anomalous and normal events & 11 \\
\cmidrule{1-3}
5 & Rare mixture of correct behavior & 21 \\
\cmidrule{1-3}
6 & Configuration change & 4 \\
\cmidrule{1-3}
7 & Manual operation & 2 \\
\bottomrule
\end{tabular}
\end{table}

\begin{figure*}[tbp]
\centering
\scriptsize
\begin{tabular}{cc}
\begin{minipage}{0.45\linewidth}
\begin{lstlisting}
258290,"2015-01-01 12:03:22","droid_movediff",NULL,"2,-3"
258291,"2015-01-01 12:03:44","droid_status",NULL,"Waiting"
258292,"2015-01-01 12:03:44","droid_tank_pos",12,NULL
258293,"2015-01-01 12:03:44","droid_lift_pos",6,NULL
258294,"2015-01-01 12:03:57","feed_tank_2",4800,NULL
258295,"2015-01-01 12:04:08","droid_swing_h",150,NULL
258296,"2015-01-01 12:04:08","droid_swing_v",150,NULL
258297,"2015-01-01 12:04:08","droid_status",NULL,"Operating"
258298,"2015-01-01 12:04:11","droid_movediff",NULL,"4,4"
258299,"2015-01-01 12:06:12","droid_lift_pos",10,NULL
258300,"2015-01-01 12:08:07","feed_tank_2",1000,NULL
\end{lstlisting}
\end{minipage}
&
\begin{minipage}{0.45\linewidth}
\begin{lstlisting}
223025,"2014-11-28 12:02:30","droid_movediff",NULL,"2,-3"
223026,"2014-11-28 12:02:59","droid_status",NULL,"Waiting"
223027,"2014-11-28 12:02:59","droid_tank_pos",12,NULL
223028,"2014-11-28 12:02:59","droid_lift_pos",6,NULL
223029,"2014-11-28 12:03:22","feed_tank_2",4800,NULL
223030,"2014-11-28 12:03:40","droid_swing_h",150,NULL
223031,"2014-11-28 12:03:41","droid_swing_v",150,NULL
223032,"2014-11-28 12:03:42","droid_status",NULL,"Operating"
223033,"2014-11-28 12:03:45","droid_movediff",NULL,"4,4"
223034,"2014-11-28 12:05:03","droid_status",NULL,"Waiting"
223035,"2014-11-28 12:05:04","droid_tank_pos",16,NULL
\end{lstlisting}
\end{minipage}
\\
(a) Outlier 1: Failure of mutual exclusion. & (b) $k$-neighborhood of outlier 1.\\

\begin{minipage}{0.45\linewidth}
\begin{lstlisting}
464104,"2015-06-28 13:20","level_3",0,NULL
464105,"2015-06-28 13:20","lightning",NULL,NULL
464106,"2015-06-28 13:20","humidity",0,NULL
464108,"2015-06-28 13:30","level_3",0,NULL
464109,"2015-06-28 13:30","lightning",NULL,NULL
464110,"2015-06-28 13:30","humidity",0,NULL
464111,"2015-06-28 13:37:41","light3_status",NULL,"on"
464112,"2015-06-28 13:37:41","light2_status",NULL,"on"
464113,"2015-06-28 13:37:41","light1_status",NULL,"on"
464114,"2015-06-28 13:37:41","light3_ontime",NULL,"06:35"
464115,"2015-06-28 13:37:41","light1_ontime",NULL,"06:45"
\end{lstlisting}
\end{minipage}
&
\begin{minipage}{0.45\linewidth}
\begin{lstlisting}
408976,"2015-05-22 17:40","air",26.2,NULL
408977,"2015-05-22 17:40","humidity",22.4,NULL
408978,"2015-05-22 17:50","pressure",1004,NULL
408979,"2015-05-22 17:50","water2",26.6,NULL
408980,"2015-05-22 17:50","level_3",9,NULL
408981,"2015-05-22 17:50","water1",24.9,NULL
408982,"2015-05-22 17:50","water3",25.8,NULL
408983,"2015-05-22 17:50","air",26.2,NULL
408984,"2015-05-22 17:50","humidity",22.4,NULL
408985,"2015-05-22 17:58:02","light1_status",NULL,"off"
408986,"2015-05-22 17:58:02","light1_ontime",NULL,"06:49"
\end{lstlisting}
\end{minipage}
\\
(c) Outlier 2: Unexpected reboot. & (d) $k$-neighborhood of outlier 2.\\

\begin{minipage}{0.45\linewidth}
\begin{lstlisting}
715630,"2016-01-02 12:03:51","feed_tank_1",4000,NULL
715631,"2016-01-02 12:04:03","droid_swing_h",0,NULL
715632,"2016-01-02 12:04:03","droid_swing_v",5,NULL
715633,"2016-01-02 12:04:03","droid_status",NULL,"Operating"
715634,"2016-01-02 12:04:06","droid_movediff",NULL,"2,0"
715635,"2016-01-02 12:04:20","droid_status",NULL,"Waiting"
715636,"2016-01-02 12:04:20","droid_tank_pos",12,NULL
715637,"2016-01-02 12:04:20","droid_lift_pos",0,NULL
715638,"2016-01-02 12:04:37","droid_swing_h",0,NULL
715639,"2016-01-02 12:04:37","droid_swing_v",-20,NULL
715640,"2016-01-02 12:04:37","droid_status",NULL,"Operating"
\end{lstlisting}
\end{minipage}
&
\begin{minipage}{0.45\linewidth}
\begin{lstlisting}
709662,"2015-12-28 12:02:43","feed_tank_1",4000,NULL
709663,"2015-12-28 12:02:55","droid_swing_h",0,NULL
709664,"2015-12-28 12:02:55","droid_swing_v",5,NULL
709665,"2015-12-28 12:02:55","droid_status",NULL,"Operating"
709666,"2015-12-28 12:02:59","droid_movediff",NULL,"2,0"
709667,"2015-12-28 12:03:12","droid_status",NULL,"Waiting"
709668,"2015-12-28 12:03:12","droid_tank_pos",12,NULL
709669,"2015-12-28 12:03:12","droid_lift_pos",0,NULL
709670,"2015-12-28 12:03:42","feed_tank_2",10000,NULL
709671,"2015-12-28 12:03:53","droid_swing_h",0,NULL
709672,"2015-12-28 12:03:53","droid_swing_v",-20,NULL
\end{lstlisting}
\end{minipage}
\\
(e) Outlier 3: Failure of single functionality. & (f) $k$-neighborhood of outlier 3.\\

\begin{minipage}{0.45\linewidth}
\begin{lstlisting}
630999,"2015-11-01 12:05:13","droid_status",NULL,"Waiting"
631000,"2015-11-01 12:05:13","droid_tank_pos",12,NULL
631001,"2015-11-01 12:05:14","droid_lift_pos",5,NULL
631002,"2015-11-01 12:05:40","feed_tank_2",10000,NULL
631003,"2015-11-01 12:05:40","target_192.168.68.93_status",NULL,"Lost"
631004,"2015-11-01 12:05:47","droid_swing_h",0,NULL
631005,"2015-11-01 12:05:48","droid_swing_v",-20,NULL
631006,"2015-11-01 12:05:48","droid_status",NULL,"Operating"
631007,"2015-11-01 12:05:51","droid_movediff",NULL,"4,2"
631008,"2015-11-01 12:06:33","droid_status",NULL,"Waiting"
631009,"2015-11-01 12:06:33","droid_tank_pos",16,NULL
\end{lstlisting}
\end{minipage}
&
\begin{minipage}{0.45\linewidth}
\begin{lstlisting}
623096,"2015-10-26 12:05:27","droid_status",NULL,"Waiting"
623097,"2015-10-26 12:05:27","droid_tank_pos",12,NULL
623098,"2015-10-26 12:05:27","droid_lift_pos",5,NULL
623099,"2015-10-26 12:05:58","feed_tank_2",10000,NULL
623100,"2015-10-26 12:06:15","droid_swing_h",0,NULL
623101,"2015-10-26 12:06:15","droid_swing_v",-20,NULL
623102,"2015-10-26 12:06:15","droid_status",NULL,"Operating"
623103,"2015-10-26 12:06:19","droid_movediff",NULL,"4,2"
623104,"2015-10-26 12:07:08","droid_status",NULL,"Waiting"
623105,"2015-10-26 12:07:08","droid_tank_pos",16,NULL
623106,"2015-10-26 12:07:08","droid_lift_pos",7,NULL
\end{lstlisting}
\end{minipage}
\\
(g) Outlier 4: Mixture of anomalous and normal events. & (h) $k$-neighborhood of outlier 4.\\
\begin{minipage}{0.45\linewidth}
\begin{lstlisting}
15191,"2014-05-17 12:02:54","feed_tank_1",300,NULL
15192,"2014-05-17 12:03:05","droid_status",NULL,"Operating"
15193,"2014-05-17 12:03:38","droid_status",NULL,"Waiting"
15194,"2014-05-17 12:03:38","droid_tank_pos",12,NULL
15195,"2014-05-17 12:03:38","droid_lift_pos",5,NULL
15196,"2014-05-17 12:03:39","fan3_status",NULL,"off"
15197,"2014-05-17 12:03:47","feed_tank_2",4000,NULL
15198,"2014-05-17 12:04:00","droid_status",NULL,"Operating"
15199,"2014-05-17 12:04:37","droid_status",NULL,"Waiting"
15200,"2014-05-17 12:04:37","droid_tank_pos",16,NULL
15201,"2014-05-17 12:04:37","droid_lift_pos",7,NULL
\end{lstlisting}
\end{minipage}
&
\begin{minipage}{0.45\linewidth}
\begin{lstlisting}
4267,"2014-05-08 12:03:35","droid_status",NULL,"Operating"
4268,"2014-05-08 12:04:11","droid_status",NULL,"Waiting"
4269,"2014-05-08 12:04:11","droid_tank_pos",12,NULL
4270,"2014-05-08 12:04:11","droid_lift_pos",5,NULL
4271,"2014-05-08 12:04:14","droid_swing_h",120,NULL
4272,"2014-05-08 12:04:14","droid_swing_v",125,NULL
4273,"2014-05-08 12:04:20","feed_tank_2",4000,NULL
4274,"2014-05-08 12:04:30","droid_status",NULL,"Operating"
4275,"2014-05-08 12:05:06","droid_status",NULL,"Waiting"
4276,"2014-05-08 12:05:06","droid_tank_pos",16,NULL
4277,"2014-05-08 12:05:06","droid_lift_pos",7,NULL
\end{lstlisting}
\end{minipage}
\\
(i) Outlier 5: Rare mixture of correct behavior. & (j) $k$-neighborhood of outlier 5.\\

\end{tabular}
\caption{Examples of outliers and their $k$-neighborhoods.}
\label{fig:ex_outliers}
\end{figure*}

\subsubsection{\bf Failure of Mutual Exclusion}

This outlier represents failures of mutual exclusion in the system.
The control for moving and feeding by the droid unit requires mutual exclusion.
However, the outlier analysis found a violation of mutual exclusion.

Figs.~\ref{fig:ex_outliers}(a) and (b) show examples of outliers and inliers, respectively, for this class.
In the outlier, the sequence after \#258298 was faulty.
The control of the moving by the droid unit must be exclusively given by a single process.
The program indicated critical sections using the ``\verb|droid_status,NULL,Operating|'' and ``\verb|droid_status,NULL,Waiting|'' entries.
Any process except one must not use the droid unit after ``\verb|droid_status,NULL,Operating|'' until ``\verb|droid_status,NULL,Waiting|'' appears.
The inlier log shows this operation between \#223032 and \#223034 in Fig.~\ref{fig:ex_outliers}(b).
In the outlier case of Fig.~\ref{fig:ex_outliers}(a), although the ``\verb|droid_status,NULL,Waiting|'' must follow after the log of ``\verb|droid_movediff|'', in the real log, the log of droid position (``\verb|droid_lift_pos,10,NULL|'') followed, and no ``\verb|droid_status,NULL,Waiting|'' log appeared after that.
The manual feeding operation for tank 2 shown in \#258300 caused this conflict.
We found that the manual feeding command during the automatic feeding process violated the mutual exclusion for control of the droid unit.

\subsubsection{\bf Unexpected Reboot}

Our target system sometimes rebooted unexpectedly.
In the outlier for this category shown in Fig.~\ref{fig:ex_outliers}(c), the log entries of \#464114 and \#464115 are important.
There were logs including \verb|light3_ontime| and \verb|light1_ontime| in the early afternoon.
These commands adjust the time to turn on the lighting the next morning.
Normally, these commands are executed at sunset, as shown in Fig.~\ref{fig:ex_outliers}(d).
However, upon rebooting, these commands are also executed to adjust the lighting time.
In this case, the anomalous executions of \verb|light3_ontime| and \verb|light1_ontime| commands imply that the system was rebooted and initialized.

\subsubsection{\bf Failure of Single Functionality}

This anomaly represents failures of a single function.
For example, missing scheduled feeding operations and lost connections of the top view camera are included in this category.
This kind of anomaly is usually found as a single log entry.

The aquarium system equips a camera that provides the top view of the aquarium.
A Raspberry Pi operates the camera server for this purpose and connects it to the aquarium system by a wireless local area network (LAN).
The wireless LAN adapter of this server sometimes lost connection, and the server restarted the LAN adapter on its own.

When the outlier (Fig.~\ref{fig:ex_outliers}(e)) and its $k$-neighborhood (Fig.~\ref{fig:ex_outliers}(f)) are compared, the entry \verb|target_X.X.X.X_status,NULL,Lost| caused the difference, where \verb|X.X.X.X| is an IP address of the camera server.
This means that the network connection of the camera server was lost and could not recover the connection by itself.

\subsubsection{\bf Mixture of Anomalous and Normal Events}

Our analysis found interesting sequences in the log.
As shown in Fig.~\ref{fig:ex_outliers}(h), many sensors successively output incorrect values to the log because both the humidity and water level were 0.
Around such sequences, the correct values can be found as outliers.
Fig.~\ref{fig:ex_outliers}(g) shows a correct sensor reading in an incorrect sensor reading sequence.
In this case, the cause of the outlier was the non-anomalous sequence of sensor commands (\#464045--\#464049), and the anomalous sequence of sensor commands in Fig.~\ref{fig:ex_outliers}(h) is recognized as an inlier sequence.

\subsubsection{\bf Rare Mixture of Correct Behavior}

Our method identified rare events during regular operations as an outlier.
Fig.~\ref{fig:ex_outliers}(i) shows an outlier in this category, and Fig.~\ref{fig:ex_outliers}(j) is an example of entries in the nearest cluster from the outlier.
They show sequences of regular automatic feeding around noon.
If we compare these sequences, we find that the \verb|fan3_status| command at \#15196 in Fig.~\ref{fig:ex_outliers}(i) caused the difference.
This means that external fan for the tank 3 starts cooling.
Because it is very rare that such unrelated commands are observed during feeding, this was detected as an outlier.

\subsubsection{\bf Configuration Change}

Changes to system settings are detected as an outlier.
Such changes are usually done to calibrate the feeding and camera positions, adjust the air conditioner target temperature, and so on.

\subsubsection{\bf Manual Operation}

Because most operations in the system are automatic, manual operation can be identified as an outlier.
For example, early feeding for a specific tank is detected as an outlier.
In this aquarium system, manual feeding rarely happens because the system allows a few feeds per day for each tank.
This is why our method detects unusual feeding commands as an outlier.

\subsection{Observation}
We expected that our method would be able to detect a single rare entry of a log, such as the failure of a single functionality.
Surprisingly, our method can also detect a complex anomaly that involves several commands, such as the failure of mutual exclusion.
This is the virtue of using a moving window rather than a single log entry to perform outlier detection.
The identification of a real unknown bug shows the capability of our method.

On the other hand, some kinds of anomalies cannot be detected as outliers by our method.
This is because, if many anomalies exist in the log and are similar (e.g., all sensor readings are 0), they form highly agglomerated clusters and thus become inliers according to our method.
Interestingly, outliers detected by our method often showed partly normal and partly anomalous behavior.
This is because they came near a cluster of anomalous values and thus were partly anomalous but differed from them by being partly normal.
Such sequences can be an entry point to anomalous behavior. Thus, finding such sequences can be useful.

Another interesting observation is that, for each chunk, the obtained outliers had quite different characteristics.
For example, outliers of chunk 3 consisted of sensor errors, while outliers of chunk 8 were related to sensing and managing the water level.
This is useful for finding problems that the target system faces in each development phase.

\section{Conclusions and Future Work}
\label{sec:CON}

We applied an outlier detection method, LOF, to find anomalies in a CPS log and evaluated the usefulness of our method by analyzing outliers it detected in an aquarium management system.
Based on our analysis results, the proposed method can detect many interesting events, such as failures of mutual exclusion, unexpected reboots, and failures of single functionality.
It also can detect manual operation of the system, which is usually harmless but can indicate a malicious attempt to control the system.

On the other hand, our method is not suitable for detecting anomalies that occur many times in similar forms; such anomalies form a cluster and are judged as inliers by our method.
Still, even in such a case, our method can detect log sequences that are partially similar to such anomalies but partially correct.
Such log sequences can be entry points to anomalous modes in the system and thus can be useful for finding the causes of anomalies.
%
% On the more technical side, we used the LOF to find outliers in 2101--3311-dimensional space.
% Although the dimension number is very high, our method successfully discerned outliers without a problem.
% This is interesting because outlier detection in high-dimensional space is usually claimed to be very challenging~\cite{Aggarwal2013}.

Future work will involve comparing the LOF with other outlier detection methods for detecting CPS anomalies.
Our method currently cannot detect anomalies that frequently occur and form agglomerated clusters.
This could be overcome by using other outlier detection methods, such as a robust principal component analysis (PCA)-based method~\cite{candes2011robust} and isolation forests~\cite{liu2008isolation}.
% To overcome this problem, we are interested in (or planning) using \emph{global} outlier detection methods, such as a robust principal component analysis (PCA)-based method~\cite{candes2011robust} or isolation forests~\cite{liu2008isolation}.

% conference papers do not normally have an appendix

% use section* for acknowledgment
\section*{Acknowledgment}

We thank Hisashi Kashima for his helpful comments.

% trigger a \newpage just before the given reference
% number - used to balance the columns on the last page
% adjust value as needed - may need to be readjusted if
% the document is modified later
%\IEEEtriggeratref{8}
% The "triggered" command can be changed if desired:
%\IEEEtriggercmd{\enlargethispage{-5in}}

% references section

% can use a bibliography generated by BibTeX as a.bbl file
% BibTeX documentation can be easily obtained at:
% http://mirror.ctan.org/biblio/bibtex/contrib/doc/
% The IEEEtran BibTeX style support page is at:
% http://www.michaelshell.org/tex/ieeetran/bibtex/
%\bibliographystyle{IEEEtran}
% argument is your BibTeX string definitions and bibliography database(s)
%\bibliography{IEEEabrv,../bib/paper}
%
% <OR> manually copy in the resultant.bbl file
% set second argument of \begin to the number of references
% (used to reserve space for the reference number labels box)
% \begin{thebibliography}{1}
%
% \bibitem{IEEEhowto:kopka}
% H.~Kopka and P.~W. Daly, \emph{A Guide to \LaTeX}, 3rd~ed.\hskip 1em plus
% 0.5em minus 0.4em\relax Harlow, England: Addison-Wesley, 1999.
%
% \end{thebibliography}

\bibliographystyle{IEEEtran}
\bibliography{IEEEabrv,IEEE.bib,LogAnalysis.bib,cps.bib,rv.bib,outlier.bib,libraries.bib}

% that's all folks
\end{document}